\documentclass[titlepage,12pt]{article}

\usepackage{amssymb,epsfig,pslatex}
\usepackage{cite}
\usepackage{amsmath}
\usepackage{graphicx}
\usepackage{float}
\usepackage{wrapfig}
\usepackage{hhline}
\usepackage{dcolumn}

\restylefloat{figure}

\newcommand{\U}{{\cal U}}

\begin{document}

\baselineskip 20pt

\vskip 20pt

\begin{center}

{\LARGE{\bf Unparticle Effects on Top Quark Pair Production at
 Photon Collider}}\\
\vspace{20pt}
{\small \it Hai-Feng Li\footnote{lihaifeng@mail.sdu.edu.cn},
Hong-lei Li,\footnote{lihl@mail.sdu.edu.cn},
Zong-Guo Si\footnote{zgsi@sdu.edu.cn},
Zhong-Juan Yang\footnote{yangzhongjuan@mail.sdu.edu.cn; Tel:13583169702}\\
Department of Physics, Shandong University, Jinan, 250100,
P. R. China}

\vspace{20pt}
{\bf Abstract}
\end{center}
\noindent The unparticle effects on $t\bar t$ production at the
future photon collider are investigated. Distributions of $t\bar t$
invariant mass and that for transverse momentum of top quark with
respect to Standard Model and unparticle production are predicted.
An odd valley with scalar unparticle contribution  appears for some
values of $d_{\U}$, which is due to the big cancellation between the
contribution from SM and that from unparticle. This character may be
used to study the properties of scalar unparticle. Our
investigations  also show  that scalar unparticle may play a
significant role in $t \bar t$ production at photon collider if it
exists.

\noindent
{\bf Key Words:} unparticle, top quark, linear collider\\
{PACS:} 14.80.-j, 14.65.Ha, 12.90.+b, 12.38.Qk

\vspace{0.5cm}

\section{Introduction}

It is believed that new physics beyond Standard Model (SM) must exist due to the
neutrino oscillation. But up to now, nobody knows exactly what the origin of
new physics is.
Though many investigations within the frame work of, for example, various
extensions of
SM Higgs doublet model, minimal supersymmetric extension of SM, technicolor,
extra dimension,
etc, have already been finished, more ideas on new physics are still necessary.
Recently, Motivated by Banks-Zaks theory \cite{Banks:1981nn}, Georgi
proposes a
fantastic idea, called unparticle physics \cite{Georgi:2007ek}. It is
a low energy effective description of a scale invariant sector
with a continuous mass distribution.
In Georgi's scheme,
 the scale invariant sector
($\cal BZ$ sector)\cite{Banks:1981nn} interacts with the SM by exchanging
particles with a very
high mass scale $M_\U^k$
\begin{equation}
   \frac{1}{M_\U^k} \,
   {\cal O}_{\mathrm SM} \, {\cal O}_{\cal BZ}  \; ,
\label{ampt}
\end{equation}
where ${\cal O}_{\mathrm SM}$ (${\cal O_{BZ}}$) represents local operator
constructed out of standard model (${\cal BZ}$ fields)
with scale dimension $d_{\mathrm SM}$ ($d_{\cal BZ}$).
Renormalization effects of the $\cal BZ$
sector induce dimensional transmutation at an energy scale
$\Lambda_\U$. Below the scale $\Lambda_\U$,
$\cal O_{\cal BZ}$ matches onto unparticle operator ${\cal O_\U}$ and
 Eq.(\ref{ampt}) becomes
\begin{equation}
 \label{effectiveop}
    C_{\cal O_\U} \frac{ \Lambda_\U^{d_{\cal BZ} - d_\U} } {M^{k}_\U } \,
   {\cal O}_{\mathrm SM}\, {\cal O}_\U \;,
\end{equation}
where $C_{\cal O_\U}$ is a coefficient function and $d_\U$ the scale
dimension of the unparticle operator. Recently, many phenomenology
studies on  unparticle have  been finished \cite{Liao:2007ic}.

On the other hand, top quark physics will play an important role in
testing the SM and finding
new physics near TeV scale. This kind of topic is widely investigated at
Tevatron, LHC and the future linear collider (LC)
 \cite{Dittmaier:1998dz}. The effects of unparticle-couplings of the SM fields
(matter and gauge) in top quark pair production at the Tevatron and
LHC are studied in ref.\cite{Choudhury:2007cq}.
It is found
in ref.\cite{Alan:2007ss} that the existence of unparticle may lead
to measurable enhancements on the SM predictions. Comparing with
$e^+ \,e^- \to  t \,\bar t$ process, $\gamma \, \gamma \to t \, \bar
t$ process will  provide information on possible anomalous $\gamma
\, t \, \bar t$ couplings
  without
contributions from $Z \, t \, \bar t $ couplings \cite{Choi:1995kp}
and some useful clues on photon photon interactions. Heavy quark
production in polarized $\gamma \, \gamma$ collisions will also help
to determine the parity of the intermediate state, e.g. Higgs boson
produced as a resonance or unparticle, etc.
In this paper, we study the unparticle effects on $t \,
\bar t$ production at the future photon collider.

The effective interaction between unparticle operators and SM field
should satisfy the SM gauge symmetry and  be Lorentz
invariant.
The economical forms of interaction between gauge bosons and unparticle are
\cite{Cheung:2007ap}
\begin{equation}
\frac{C_K}{\Lambda_{\U}^{d_{\U}}}\,G_{\alpha\,\beta}\,G^{\alpha\,\beta}
O_{\U},~~~~~~~~~~~~~~~
\frac{C_K}{\Lambda_{\U}^{d_{\U}}}\,G_{\alpha\,\mu}\,G^{\alpha}_{\nu}
O_{\U}^{\mu\,\nu}
\end{equation}
where $G_{\alpha\,\beta}$ denotes the gauge boson field, $C_K$ the  gauge boson
coupling of
scalar unparticle. 
The vector and tensor unparticle operators are  Hermitian and
transverse \cite{Cheung:2007ap},
\begin{equation}
\partial_{\mu} O^\mu_{\U}=0, ~~~~~~~~~~~~~~~~
\partial_{\mu} O^{\mu \nu}_{\U}=0
\end{equation}
Additionally, the tensor unparticle operator is assumed to be traceless 
$ O^\mu_{\U \mu}=0$. Using these interaction Hamiltonians, we 
can obtain Feynman rules
for the unparticle and gauge boson interaction. There is no vector
unparticle interacting with gauge boson due to the Lorentz invariant
and transverse condition of unparticle. The corresponding Feynman
rules used in this paper are listed as follows\cite{Cheung:2007ap}:

{\bf Unparticle propagators:}
\begin{eqnarray*}
\Delta_F(P^2) &=&  \frac{i\,\,A_{d_\U}} { 2 \sin (d_\U \pi) }
(-P^2)^{d_\U -2}, \hbox{~~~~~~~~~~~~~~~~~~~~~for ~~Scalar ~unparticle, }
\nonumber\\
 \left[ \Delta_F (P^2) \right]_{\mu\nu, \rho\sigma} &=&
         \frac{i\,\,A_{d_\U}}{2 \sin (d_\U \pi)} \, (-P^2)^{d_\U -2} \,
         T_{\mu\nu, \rho\sigma}(P) \;,\hbox{~~~~~for ~~Tensor ~unparticle},
\end{eqnarray*}
with $P$ the four momentum of the unparticle, and
\begin{eqnarray*}
\label{branchcut}
 (-P^2)^{d_\U -2}&=&\left \{
\begin{array}{lcl}
|P^2|^{d_\U -2}   & \quad & \hbox{for ~$P^2$ ~is negative and real, } \\
|P^2|^{d_\U -2} e^{-i d_\U \pi} & & \hbox{for ~$P^2$~ is positive}.
\end{array} \right. \nonumber \\
   A_{d_\U}&=&{16\pi^2\sqrt{\pi}\over (2\pi)^{2{d_\U}}}
  {\Gamma({d_\U}+{1\over2})\over\Gamma({d_\U}-1)\Gamma(2\,{d_\U})},\nonumber \\
T^{\mu\nu,\rho\sigma}(P) &=& \frac{1}{2} \, \left\{
   \pi^{\mu\rho}(P)\  \pi^{\nu\sigma}(P)
        + \pi^{\mu\sigma}(P)\  \pi^{\nu\rho}(P) - \frac{2}{3}\
          \pi^{\mu\nu}(P)\  \pi^{\rho\sigma}(P)  \right\} \;,\\
\pi^{\mu\nu}(P) &=& - g^{\mu \nu} + \frac{P^\mu P^\nu }{ P^2} \;.
\end{eqnarray*}

{\bf Vertices w.r.t. scalar unparticle}
\begin{itemize}
 \item $
i\,\frac{C_S}{\Lambda_\U^{d_{\U}-1}}-\frac{C_P}
{\Lambda_\U^{d_{\U}-1}}\,
\gamma^5
+\frac{C_V }{\Lambda_\U^{d_{\U}}}\not{\!}P$ \;\;\;\;\;\;\;\;\;\; for
\,\,$t\bar t \U$
\item $4\,i\,\frac{C_K}{\Lambda_\U^{d_{\U}}}(-p_1\cdot
p_2\,g^{\mu\nu}+p_1^{\nu}p_2^{\mu})$
 \;\;\;\;\;\;\;\;\;\; for \,\, $\gamma\gamma\U$
\end{itemize}
where $C_S$, $C_P$ and $C_V$ respectively denote scalar, pseudoscalar
and vector couplings of scalar unparticle,
$p_1$ ($p_2$) the momentum of the corresponding photon, and $\U$ the
unparticle.  The vertex
$ \frac{C_V}{\Lambda_\U^{d_{\U}}}\not{\!}P$ does not contribute
to the process because the final state  fermion current is on-shell.

{\bf Vertices w.r.t. tensor unparticle}
\begin{itemize}
 \item
$-\frac{i\,\lambda_{T}}{4\,\Lambda_\U^{d_{\U}}}[\gamma^{\mu}(p_1^{\nu}-p_2^{\nu})+
\gamma^{\nu}(p_1^{\mu}-p_2^{\mu})]  $\;\;\;\;\;\;\;\;\;\; for \,\,$t\bar t\U$
\item
$i\,\frac{\lambda_{T}}{\Lambda_\U^{d_{\U}}}[K^{\mu\nu\rho\sigma}+K^{\mu\nu
\sigma\rho}
]$
\;\;\;\;\;\;\;\;\;\; \;\;\;\;\;\;\;for \,\, $\gamma\gamma\U$
\end{itemize}
with $K^{\mu\nu\rho\sigma} = -g^{\mu\nu} p_1^\rho p_2^\sigma - p_1 \cdot p_2
g^{\rho\mu} g^{\sigma \nu} + p_1^\nu p_2^\rho g^{\sigma \mu}
                           + p_2^\mu p_1^\rho g^{\sigma \nu}$.
$\lambda_T$ is the dimensionless effective coupling constant.

This paper is organized as follows: the analytical results for $t
\bar t$ production in $\gamma$ $\gamma$ collisions related to SM and
unparticle process are listed in Sec.2, and the
numerical results are shown in Sec.3. Finally we give a brief
summary.

\vspace{0.3cm}

\section{$t \bar t $ Production in $\gamma \gamma$ Collisions}
\label{sec2}

\begin{figure}
\begin{center}
\psfig{figure=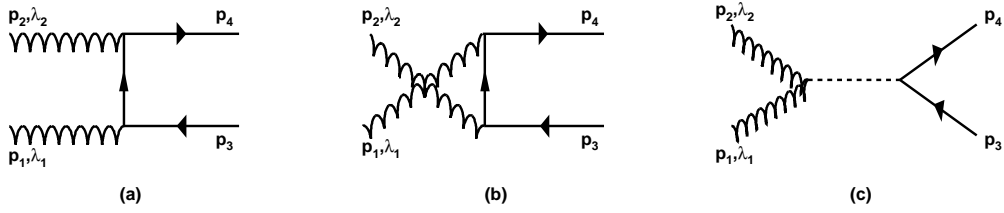,width=16cm}
\caption{ Feynman diagrams for $t \bar t$ production.
(a) and (b): SM contribution. (c): via virtual unparticle propagator.}
\label{efwtg}
\end{center}
\end{figure}

At a photon collider, top quark pair can be produced within
SM (Fig.\ref{efwtg}(a), (b))

\begin{equation}
\label{wew}
\gamma(p_1,\lambda_1)+\gamma(p_2,\lambda_2) \xrightarrow{SM}
\bar{t}(p_3)+t(p_4),
\end{equation}
or via Non-SM
particle, e.g. unparticle (Fig.\ref{efwtg}(c))
\begin{equation}
\label{wyd}
\gamma(p_1,\lambda_1)+\gamma(p_2,\lambda_2) \xrightarrow{\cal U}
\bar{t}(p_3)+t(p_4),
\end{equation}

where $p_i$ ($i=1,2,3,4$) respectively denote the momenta of the
corresponding particles,  $\lambda_1$ ($\lambda_2$) the initial photon helicity.
The differential cross section for $t\bar{t}$ pair production can be
written as:
\begin{equation}
d\hat \sigma(\lambda_1,\lambda_2)={1\over 2 \hat s} {|{\cal M}|}^2
d\Gamma_2,
\end{equation}
where  $\hat s=(p_1+p_2)^2$ and $d\Gamma_2$ the two-particle phase
space. The invariant amplitude for $t \bar t$ production at photon
collider $\cal M=\cal M_{SM}+\cal M_{\U}$ with $\cal M_{SM}$ for
process (\ref{wew}) and $\cal M_{\U}$ for process (\ref{wyd}). Then
we can obtain the matrix element square as follows:

{\bf Scalar unparticle}

\begin{eqnarray}
\label{wen2} {|{\cal M}|}^2&=&N_c \Big[ \frac{C_K}{\Lambda_\U^{2d_\U
- 1} } \,\frac{A_{d_\U}} {\sin(d_\U \pi)} \Big]^2 \,\hat s^{2d_\U-1}
(1+\lambda_1\lambda_2)({C_P}^2 +\beta^2 {C_S}^2)
\nonumber\\
&+&\frac{32\,N_c\, \pi \alpha A_{d_\U} Q_t^2
{C_K}}{\Lambda_\U^{2d_\U - 1} }
 \frac{\hat s^{d_\U-1}m}{(\beta^2z^2-1)} \Big[
{C_P}(\lambda_1+\lambda_2)-\beta^2{C_S}(1+\lambda_1\lambda_2)cot(d_\U\pi)
\Big] \nonumber\\
&+&{|{\cal M}_{SM}|}^2,
\end{eqnarray}
with $N_c=3$, the number of colors, and

\begin{eqnarray}
{|{\cal M}_{SM}|}^2&=&{64 N_c \alpha^2Q_t^4\pi^2\over
(1-\beta^2z^2)^2} \Big \{1+2\beta^2(1-z^2)-\beta^4\left[1+(1-z^2)^2
\right] \nonumber \\
&+&\lambda_1\lambda_2
\left[
1-2\beta^2(1-z^2)-\beta^4z^2(2-z^2)
\right]\Big\},
\end{eqnarray}
where $\alpha$ is the fine structure constant, $Q_t=2/3$, $m$ the top quark
mass, $\beta=\sqrt{1-{4m^2/\hat s}}$,
and $z=\hat{\bf p_1}\cdot\hat{\bf p_4}$, with $\hat{\bf p_1}$
($\hat{\bf p_4}$) the direction of the initial photon
(top quark) in the $\gamma $ $\gamma$ Center of Mass System (CMS).
${|{\cal M}_{SM}|}^2$ agrees with that in ref.\cite{Brandenburg:2005uu}.

{\bf Tensor  unparticle}

\begin{eqnarray}
\label{wen3} {|{\cal M}|}^2&=&N_c\, \Big[\frac{
\lambda_T^2}{8\Lambda_\U^{2d_\U} }\,\frac{A_{d_\U}} { \sin (d_\U
\pi) }\Big]^2\,\hat s^{2d_{\U}}\Big\{
\beta^2(\lambda_1\lambda_2-1)(z^2-1)\left[2+\beta^2(z^2-1)\right]
\Big\} \nonumber \\
&+&\frac{ 2 \pi N_c \alpha A_{d_\U} Q_t^2\lambda_T^2
}{\Lambda_\U^{2d_\U} } \frac{\hat s^{d_\U}}{(\beta^2z^2-1)} \Big\{
\beta^2(\lambda_1\lambda_2-1)(z^2-1)\left[2+\beta^2(z^2-1)
\right]cot(d_\U\pi)\Big\}\nonumber \\
&+& {|{\cal M}_{SM}|}^2
\end{eqnarray}
The terms in the second line of Eq.(\ref{wen2}) and (\ref{wen3}) are
the interference terms between SM and unparticle process.

\section{Numerical results  for $\gamma \gamma \to t \bar t $ }
\label{sec3}

The future linear collider is a large scale project in
accelerator particle physics with electron-positron colliding at
energies from 0.5 TeV up to about 1 TeV \cite{Phinney:2007zz}.
LC can be operated in photon
photon mode, where high energy photons can be obtained by Compton
backscattering of laser light off the high energy electron beam
\cite{Ginzburg:1982yr}.
Combining the results in Sec.2, we obtain the effective cross section for
top quark pair production at the LC
\begin{equation}
\label{eeq14}
\sigma (S,P_{e1},P_{e2},P_{L1},P_{L2})=\int_{0}^{y_{max}} dy_1 \int_{0}^{y_{max}}
dy_2 f_{\gamma}^e
(y_1,P_{e1},P_{L1}) f_{\gamma}^e(y_2,P_{e2},P_{L2}) \hat{\sigma}
(\hat s,\lambda_1,\lambda_2)
\end{equation}
The function $f_{\gamma}^e(y,P_e,P_L)$ is the normalized energy spectrum of
the photons
\begin{equation}
\label{rre}
f_{\gamma}^e(y,P_e,P_L)={\cal N}^{-1}[\frac{1}{1-y}-y+(2r-1)^2
-P_eP_Lxr(2r-1)(2-y)],
\end{equation}
where $ \sqrt{S}$ is the $e^+e^- $ CMS energy, $\cal N$ the normalization
factor, $P_e $ $(P_L)$ the polarization of
the initial electron (laser) beam,
$r=y/(x-xy)$, and $y$ is the fraction of the electron energy
transferred to the photon in the center of mass frame. It has the following
range
\begin{equation}
0\leq y \leq   y_{max}\equiv\frac{x}{x+1}, \hspace {1.5cm}
 x=\frac{4 E_L E_e}{m_e^2},
\end{equation}
with $m_e$ is the electron mass and $E_e$ ($E_L$) the energy of the
electron (laser) beam. In order to avoid the creation of an $e^+e^-$
pair from the backscattered photon  and the initial photon,
one has to set $x \leq 2(1+\sqrt{2})$.
$\lambda_i (i=1,2)$ presented in the cross section
$ \hat{\sigma}(\hat s,\lambda_1,\lambda_2)$
is now given by
\begin{equation}
\lambda_i=P_\gamma(y_i,P_{e(i)},P_{L(i)}), i=1,2,
\end{equation}
where the function $P_\gamma(y,P_e,P_L)$ is the polarization
of photons scattered with energy fraction $y$,
\begin{equation}
P_\gamma(y,P_e,P_L)=\frac{1}{f_{\gamma}^e(y,P_e,P_L)\cal N}
\{xrP_e[1+(1-y)(2r-1)^2]-(2r-1)P_L
[\frac{1}{1-y}+1-y]\}.
\end{equation}

\begin{figure}
\begin{center}
 \psfig{figure=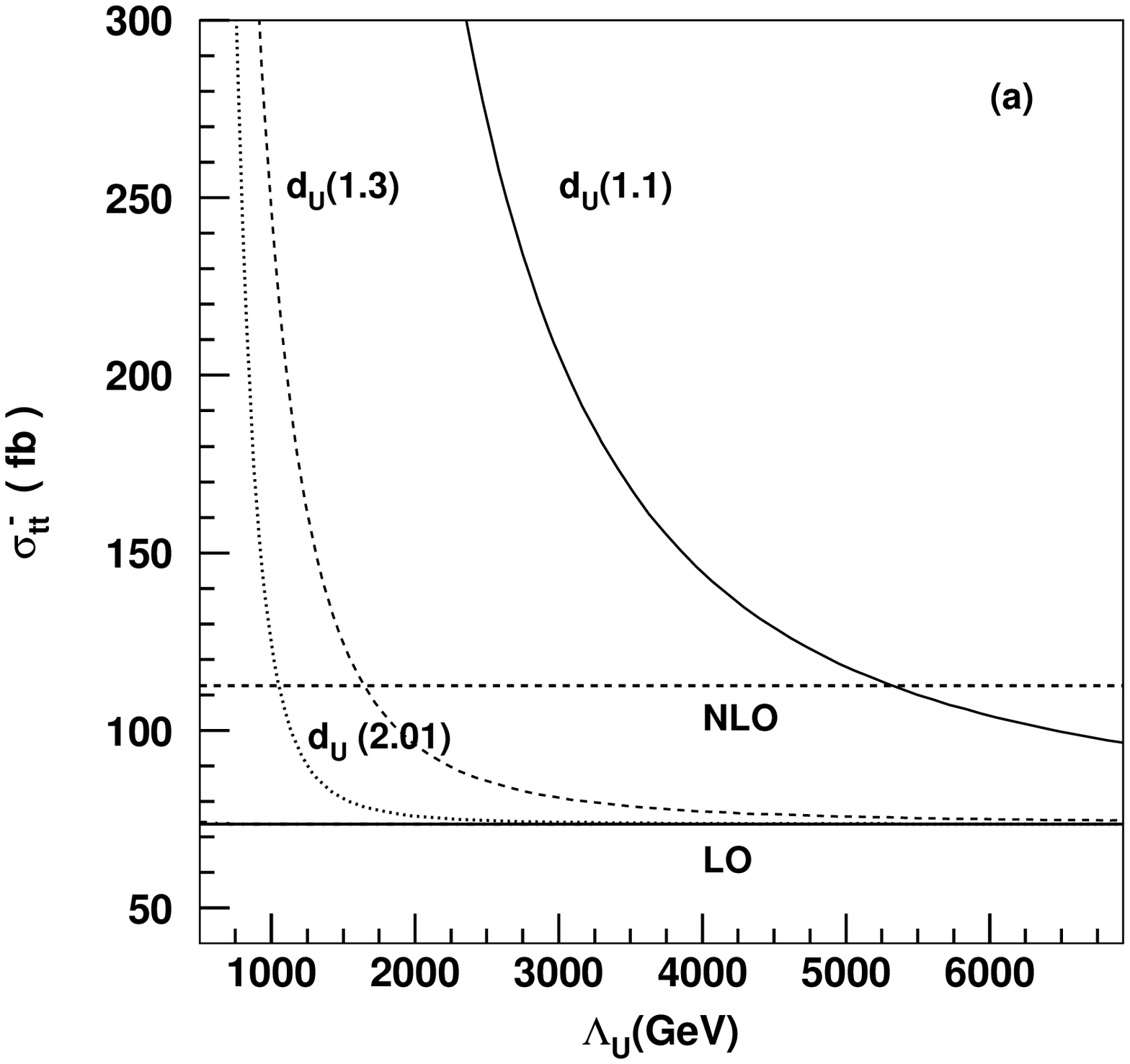,width=8cm}  \psfig{figure=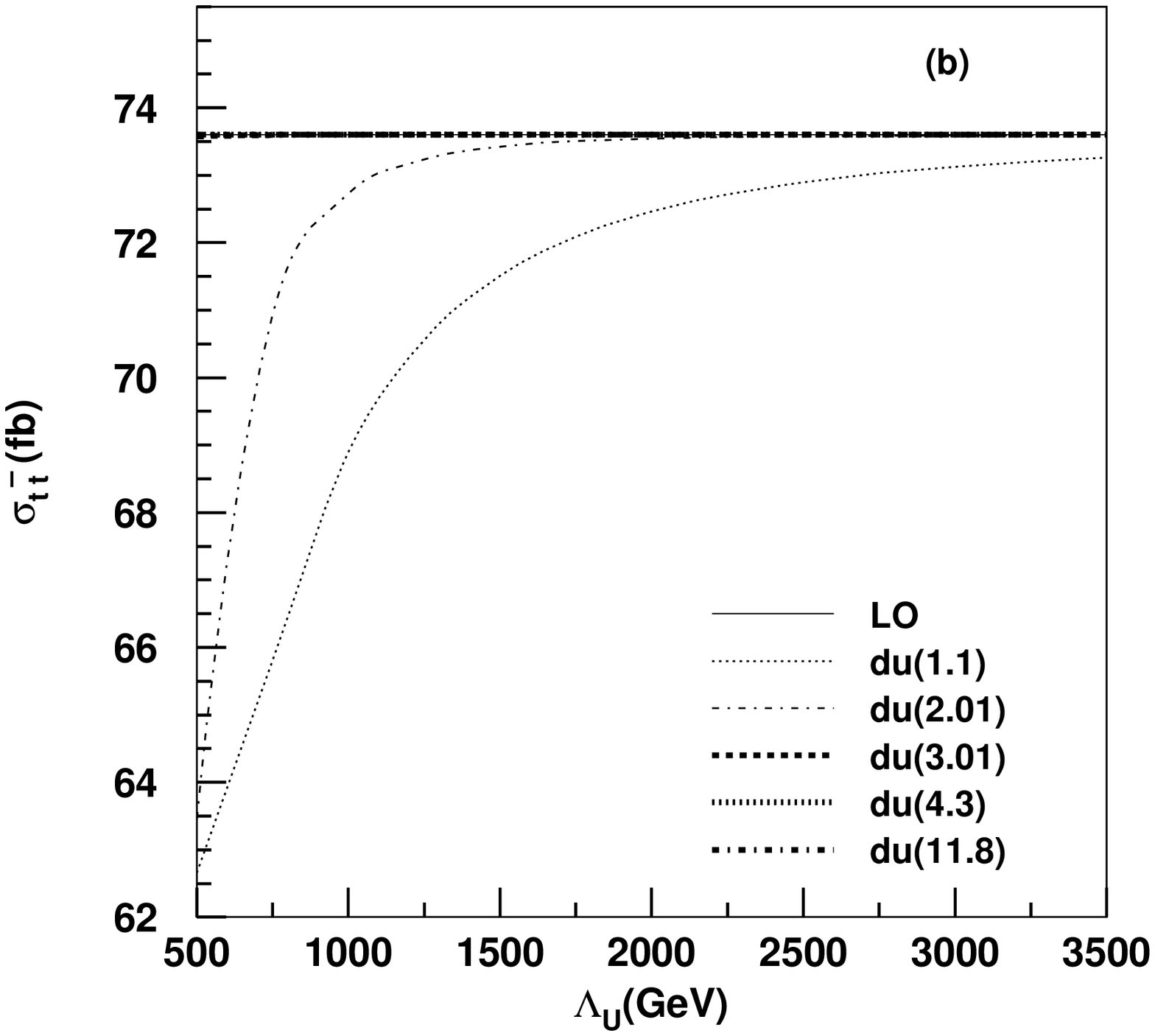,
width=8cm}
\caption{The total cross section for $t \bar t$ production at photon collider
(a) unpolarized beams with scalar unparticle. (b) unpolarized
beams with tensor unparticle.}
\label{tevatron}
\end{center}
\end{figure}

In our numerical calculation, we adopt the inputs
$E_e=250 GeV$, $E_L= 1.26 eV$, $m_e=5.11\times10^{-4} GeV$,
$m_t=172.5 GeV$ and $\alpha=1/128$.
In order to give a naive estimation at LC, we set
\begin{equation}
  C_K= C_P=C_S \equiv \lambda_S,
\end{equation}
where $\lambda_S$ denotes the dimensionless effective coupling constant.
Top quark pair production with unparticle effects at Tevatron is first
calculated in \cite{Choudhury:2007cq},
where the constraints on the unparticle energy scale $\Lambda_{\U}$ are
obtained. When $d_{\U}=1.1 (2.01)$ and
$\Lambda_{\U}>600  (1200)$ GeV,
the total cross section for $t \bar t$ production at Tevatron given
in \cite{Choudhury:2007cq} is within
the 95\% C.L. upper
limit of CDF experiment data.
Using the same parameters as  \cite{Choudhury:2007cq}, we give the
total cross section for $t \bar t$ production at photon collider in
fig.\ref{tevatron}. The leading order (LO) and
next-to-leading order (NLO) SM predictions are also shown  in the same figure.
Assuming the new physics effects are not far away from SM predictions and
demanding the total cross section for $t\bar t$ production is between
LO and NLO  QCD prediction,
we can obtain some constraints on $\Lambda_{\U}$ with
scalar unparticle in fig.\ref{tevatron}(a).
Our results indicate that $\Lambda_{\U}$ should be larger than 5500 (1000) GeV
when  $d_{\U}=1.1$ (2.01). This is consistent with the results in ref.
\cite{Choudhury:2007cq}.
The total cross
section with tensor unparticle in fig.\ref{tevatron}(b)
is small for $\Lambda_{\U}>1.5$ TeV. We also investigate the cross
section w.r.t. tensor unparticle for different $d_{\U}$. Obviously, 
for $d_{\U} > 3 $, the   
tensor unparticle effects can almost be neglected.

To illustrate the unparticle effects on $t \bar t$ production,
we define the following dimensionless variables
\begin{eqnarray}
R_{\U}^{S(T)}= \frac{\sigma^{S(T)}_{SM+{\U}}-\sigma_{SM}}
{\sigma_{SM}},\hspace{1 cm}
 R^{S(T)}=\big {|} \frac{\frac{d\sigma_{SM+\U}^{S(T)}}{dM_{t\bar t}}-
\frac{d\sigma_{SM}}
{dM_{t\bar t}}}{\frac{d\sigma_{SM}}{dM_{t\bar t}}} \big{|},\hspace{1 cm}
 r^{S(T)}=\big {|}\frac{\frac{d\sigma_{SM+\U}^{S(T)}}{dP_{t}}-
\frac{d\sigma_{SM}}{dP_{t}}}
{\frac{d\sigma_{SM}}{dP_{t}}}\big{|},
\end{eqnarray}
where $\sigma_{SM}$ denotes the effective cross section within SM,
$\sigma_{SM+{\U}}^{S(T)}$
the cross section of SM plus scalar (tensor) unparticle contribution,
$M_{t \bar t}$
the $t \bar t$ invariant mass, and $P_t$ the transverse momentum
of top quark.

\begin{figure}
\begin{center}
 \psfig{figure=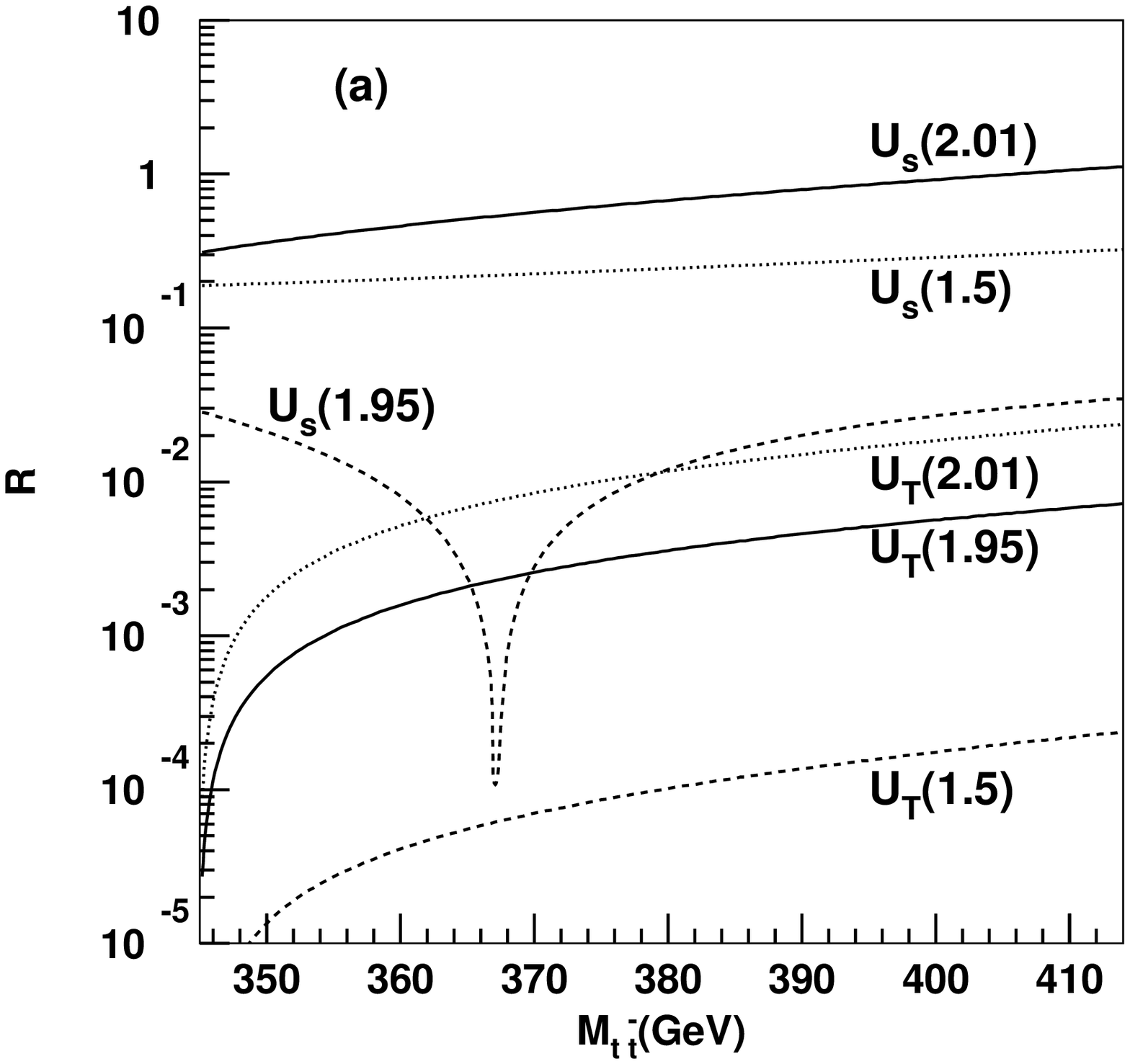,width=8cm} \psfig{figure=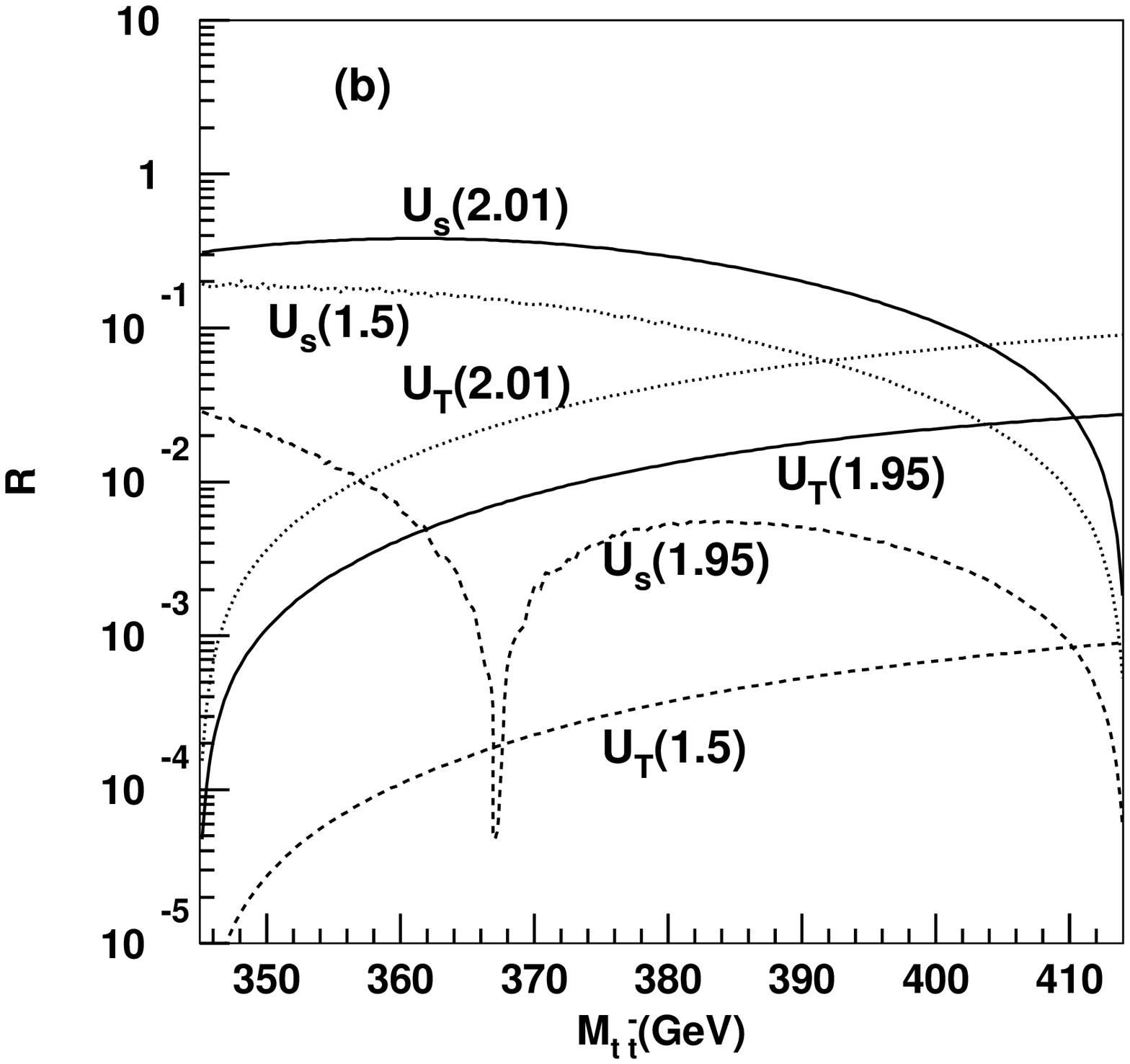,width=8cm}
\caption{$R^{S(T)}$-distribution for (a) unpolarized beams and (b)
polarized beams with
$(P_{e1},P_{e2};P_{L1},P_{L2})=(0.85,-0.85;-1,+1)$.}
\label{dyzuxw}
\end{center}
\end{figure}

\begin{figure}
\begin{center}
 \psfig{figure=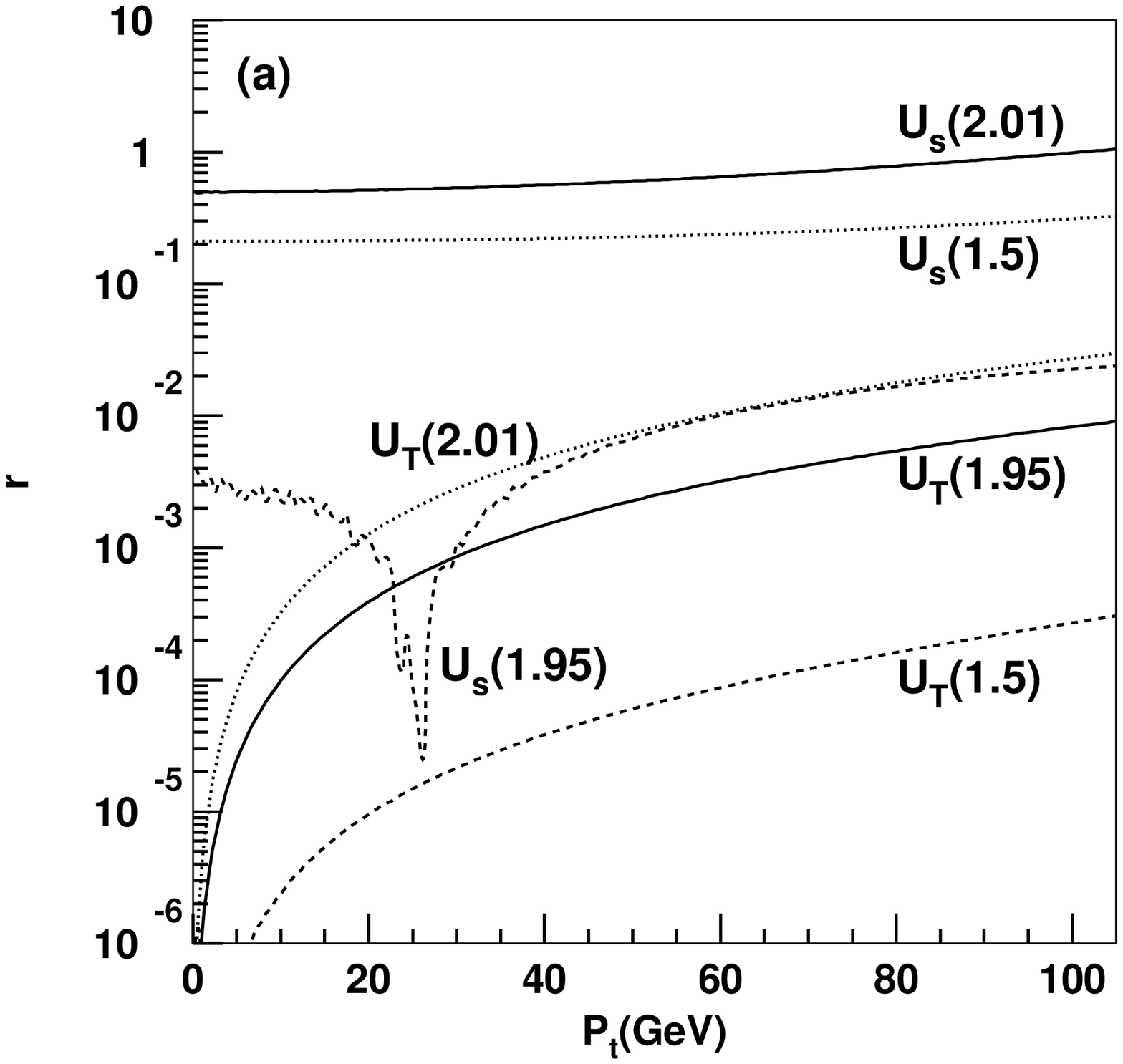 ,width=8cm} \psfig{figure=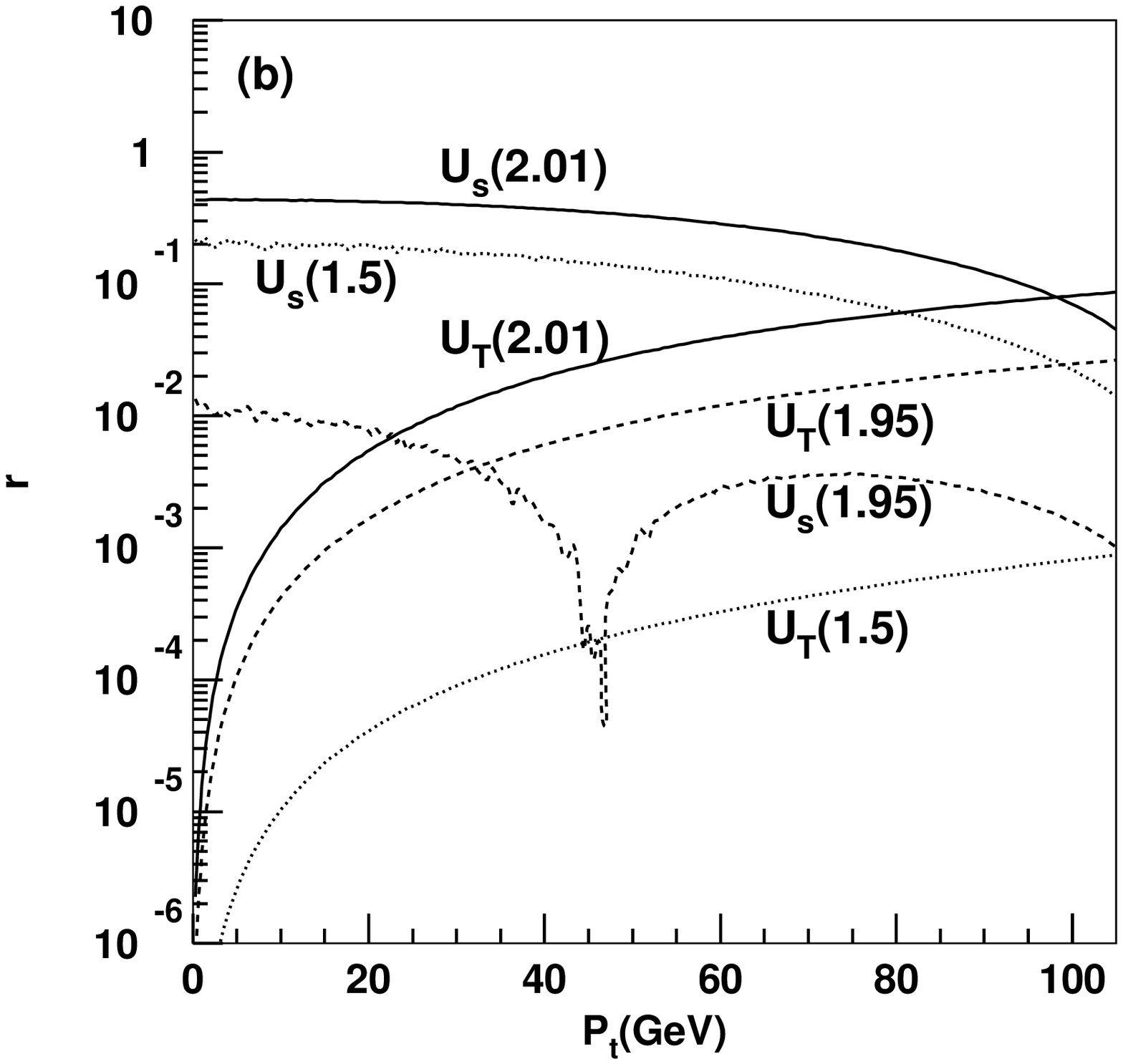,
width=8cm}
\caption{$r^{S(T)}$-distribution for (a) unpolarized beams and (b)
polarized beams with
$(P_{e1},P_{e2};P_{L1},P_{L2})=(0.85,-0.85;-1,+1)$.}
\label{yzuxqrw}
\end{center}
\end{figure}

\begin{figure}
\begin{center}
 \psfig{figure=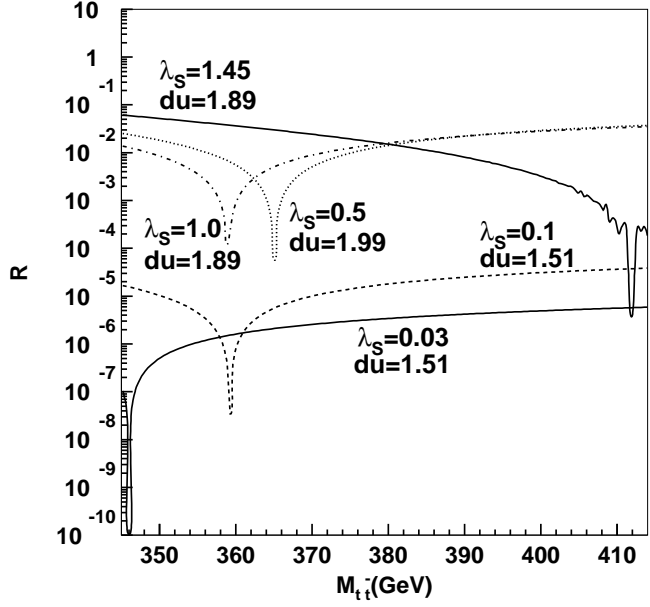 ,width=10cm}
\caption{$R^{S}$-distribution for unpolarized beams which a valley appears.}
\label{yzv}
\end{center}
\end{figure}

In Table \ref{corr1}, we provide our numerical results for the
effective cross section within SM\footnote{Our results agree to
those obtained in ref.\cite{Brandenburg:2005uu}.} and
$R_{\U}^{S(T)}$ with $\lambda_{S}=\lambda_{T}=1$, and
$\Lambda_{\U}=1TeV$.
When $d_{\U}$ is close to 2, $R_{\U}^{S(T)}$ becomes slightly large
due to the singularities of the propagator\footnote{we use the propagators for 
a scale invariant sector which differs from a conformal invariant hidden ones in
ref.\cite{Grinstein:2008qk}. When using the propagators in 
ref.\cite{Grinstein:2008qk}, there is no singularity for integer scaling 
dimensions.}. The unparticle effects
induced by individual scalar operator are given in Table
\ref{corr2}. It is found that
the unparticle effects on $t \bar t$ production depend on its spin
and scale dimension. The $M_{t\bar{t}}$ ($P_t$) distributions
$R^{S(T)}$ ($r^{S(T)}$) for unpolarized and polarized beams are
respectively shown in fig.\ref{dyzuxw}(\ref{yzuxqrw}), where
$U_{S(T)}(1.5)$ denotes the scalar (tensor) unparticle
contributions with $d_{\U}=1.5$, etc.
Our results show that the effects of scalar unparticle are
significant at unpolarized (polarized) photon collider for $M_{t
\bar t} \geq 370 GeV$ ($M_{t \bar t} \leq 362 GeV$) and $P_t \geq 30
GeV$ ($P_t \leq 25 GeV$). One interesting phenomena is that for the
scalar unparticle, a valley appears for $R^S$- ($r^S$-) distribution
at $M_{t\bar t} \approx 368 GeV$ ($P_t \approx 23 GeV$). This effect
is due to the interference between unparticle and SM contributions.
We find when $ 0.02\leq \lambda_S \leq 1.46$, a valley appears at
some values of $d_\U$ (fig.\ref{yzv}). This character may be used to
determine the properties of scalar unparticle if it exists. Finally,
the dependence of $t\bar{t}$ production induced by unparticle
effects on the couplings is investigated (Table \ref{corr3}). One can
find that $R_{\U}^{S(T)}$ becomes larger as the value of the
coupling increases.

\begin{table}
\begin{center}
\begin{tabular}{|l||l|l|l|l|l|l|l|l|l|}  \hline
 {\em $(P_{e1},P_{e2};P_{L1},P_{L2})$}& \em   ${\sigma}_{SM}(fb)$
& \multicolumn{2}{c|} {\em$d_\U=1.3$}  &  \multicolumn{2}{c|}
{\em$d_\U=1.5$} &  \multicolumn{2}{|c|} {\em$d_\U=1.95$}
&  \multicolumn{2}{|c|} {\em$d_\U=2.01$} \\
\cline{3-4} \cline{5-6} \cline{7-8} \cline{9-10}
 & & $R_{\U}^{S}$& $ R_{\U}^{T}$&$R_{\U}^{S}$& $ R_{\U}^{T}$&$R_{\U}^{S}$
&  $ R_{\U}^{T}$& $R_{\U}^{S}$& $ R_{\U}^{T}$\\  \hline  \hline
$(0,0;0,0)$& 73.6 & 2.3 &-0.01 &0.2 &0.0001 &-0.01 &0.004 &0.7 &-0.01
  \\  \hline
$(0.85,0.85;-1,-1)$&228.6&0.8 &-0.001  &-0.6 & 0.00001& -0.08&0.0004
&0.9 &-0.001 \\\hline
$(0.85,0.85;+1,+1)$&29.0& 0.7 &-0.008  &-0.4 &0.00006&-0.04
& 0.002& 0.6&-0.007 \\  \hline
$(0.85,-0.85;-1,+1)$&78.6 & 0.8 &-0.06 & 0.09& 0.0005 &-0.001
& 0.02&0.2 & -0.05 \\ \hline
\end{tabular}
\caption{Results for the effective cross section within SM and
$R_{\U}^{S(T)}$( $\lambda_{S}=\lambda_{T}=1$)
 at $\sqrt {S}=500 GeV$.}
\label{corr1}
\end{center}
\end{table}

\begin{table}
\begin{center}
\begin{tabular}{|c||c|c|c|c|c|c|c|c|}  \hline
 & \multicolumn{8}{c|} {\em$R_{\U}^{S}$}   \\ \cline{2-9}
{\em $(P_{e1},P_{e2};P_{L1},P_{L2})$}
& \multicolumn{2}{c|} {\em$d_\U=1.3$}  &  \multicolumn{2}{c|}
{\em$d_\U=1.5$} &  \multicolumn{2}{|c|} {\em$d_\U=1.95$}
&  \multicolumn{2}{|c|} {\em$d_\U=2.01$}  \\
\cline{2-3} \cline{4-5} \cline{6-7} \cline{8-9}

    & $C_S=1$ &$C_S=0$ &  $C_S=1$ & $C_S=0$ & $C_S=1$  & $C_S=0$
& $C_S=1$  & $C_S=0$ \\
  & $C_P=0$ &$C_P=1$ & $C_P=0$ & $C_P=1$& $C_P=0$ & $C_P=1$
& $C_P=0$ & $C_P=1$\\ \hline \hline
$(0,0;0,0)$ & 0.5  & 1.8 & 0.04 & 0.2 & -0.05& 0.04
& 0.3  & 0.4 \\ \hline
$(0.85,0.85;-1,-1) $& 0.8  &0.05   &0.05  & -0.7 &-0.07 & -0.01
& 0.4 &0.5  \\ \hline
$(0.85,0.85;+1,+1) $ &0.4  &0.3  &0.03  & -0.4 & -0.04 & -0.006
&0.2 & 0.4 \\ \hline
$(0.85,-0.85;-1,+1)$ & 0.2 &0.7  & 0.01& 0.08  & -0.01 &0.01
& 0.08& 0.1 \\ \hline
\end{tabular}
\caption{Results for 
$R_{\U}^{S}$($C_K=1$) with different values of
$C_S$ ($C_P$) at $\sqrt {S}=500 GeV$.}
\label{corr2}
\end{center}
\end{table}

\begin{table}
\begin{center}
\begin{tabular}{|l||l|l|l||l|l|l|l|}  \hline
& \multicolumn{3}{c||}{ $R_{\U}^{S}$} & \multicolumn{3}{c|}{$ R_{\U}^{T}$} \\
\cline{2-4}  \cline{5-7}
 & $\lambda_S=0.1$ & $\lambda_S=0.3$ & $\lambda_S=1$
& $\lambda_T=0.5$ & $\lambda_T=1$ &  $\lambda_T=3$
\\\hline  \hline
\em $d_\U=1.1$& 0.013& 0.27 &22.0 & -0.017&-0.064&0.058
\\  \hline \em $d_\U=1.5$& 0.00002&0.002  &0.247 &0.000007
&0.0001 &0.0087
\\  \hline
\em $d_\U=1.9$&-0.0004& -0.0032&-0.015  &0.0006  &0.0006
&0.023\\
\hline \em $d_\U=2.01$&0.0019   & 0.021   &0.688 &-0.003 &-0.012
&-0.091\\  \hline \em $d_\U=2.2$&0.00002  & 0.00019 &0.002 &-0.00003
&-0.0001   &-0.001\\  \hline
\em$d_\U=2.8$&-0.0000002&-0.000002&-0.00002 & 0.0000004 &0.000001
&0.00001\\  \hline
\end{tabular}
\caption{Results for $R_{\U}^{S(T)}$  with different coupling
constants
 at $\sqrt {S}=500 GeV$.}
\label{corr3}
\end{center}
\end{table}

\section{Summary}

Within SM, $t\bar t$ can be produced via u- or t-channel at photon
collider.  While unparticle  can induce top quark pair production
via s-channel. In this paper,  we investigate  $t\bar t$ production
including unparticle effects at photon collider. We find that the
unparticle effects  depend on the unparticle spin and its scale
dimension $d_{\U}$. 
We investigate the dependence of the total cross section
on $\Lambda_{\U}$. If we assume that the predictions including
unparticle effects are not far away from those obtained within SM,
we can get some constraints on $\Lambda_{\U}$. Our results show that
scalar unparticle may play a significant role in $t \bar t$
production at photon collider. The $M_{t\bar{t}}$ ($P_T$)
distributions related to scalar unparticle show that a valley exists
when $ 0.02 \leq \lambda_S \leq 1.46$ for some values of $d_\U$,
which may be used to investigate the properties of scalar
unparticle. Once the photon collider is available, it will become
possible to investigate the unparticle physics or other unexpected
new physics beyond SM.

\section*{Acknowledgments}
This work is supported in part by NSFC, NCET and Huo YingDong Foundation.
The authors thank all of the members
in  Theoretical Particle Physics Group of Shandong
University for their helpful discussions.


\begin{thebibliography}{99}



\bibitem{Banks:1981nn}
  T.~Banks and A.~Zaks,
  Nucl.\ Phys.\  B {\bf 196}, 189 (1982).

\bibitem{Georgi:2007ek}
  H.~Georgi,
  Phys.\ Rev.\ Lett.\  {\bf 98}, 221601 (2007);
  Phys.\ Lett.\  B {\bf 650}, 275 (2007).

\bibitem{Liao:2007ic}
  Y.~Liao and J.~Y.~Liu,
  Phys.\ Rev.\ Lett.\  {\bf 99}, 191804 (2007);
  K.~Cheung, W.~Y.~Keung and T.~C.~Yuan,
  Phys.\ Rev.\ Lett.\  {\bf 99} (2007) 051803;
  M.~Luo and G.~Zhu,
  Phys.\ Lett.\  B {\bf 659} (2008) 341;
  C.~H.~Chen and C.~Q.~Geng,
  Phys.\ Rev.\  D {\bf 76} (2007) 036007;
  T.~M.~Aliev, A.~S.~Cornell and N.~Gaur,
  Phys.\ Lett.\  B {\bf 657} (2007) 77;
  R.~Mohanta and A.~K.~Giri,
  Phys.\ Rev.\  D {\bf 76} (2007) 075015;
  G.~J.~Ding and M.~L.~Yan,
  Phys.\ Rev.\  D {\bf 76}, 075005 (2007);
  C.~D.~Lu, W.~Wang and Y.~M.~Wang,
  Phys.\ Rev.\  D {\bf 76}, 077701 (2007);
  H.~Davoudiasl,
  Phys.\ Rev.\ Lett.\  {\bf 99}, 141301 (2007);
  P.~Mathews and V.~Ravindran,
  Phys.\ Lett.\  B {\bf 657}, 198 (2007);
  H.~Zhang, C.~S.~Li and Z.~Li,
  Phys.\ Rev.\  D {\bf 76}, 116003 (2007);
  R.~Mohanta and A.~K.~Giri,
  Phys.\ Lett.\  B {\bf 660}, 376 (2008);
  H.~Zhang, C.~S.~Li and Z.~Li,
  Phys.\ Rev.\  D {\bf 76} (2007) 116003;
  O.~Cakir and K.~O.~Ozansoy,
  arXiv:0712.3814 [hep-ph];
  X.~Q.~Li and Z.~T.~Wei,
  Phys.\ Lett.\  B {\bf 651} (2007) 380;
  A.~Lenz,
  Phys.\ Rev.\  D {\bf 76} (2007) 065006;
  K.~Huitu and S.~K.~Rai,
  Phys.\ Rev.\  D {\bf 77}, 035015 (2008);
  T.~Kikuchi, N.~Okada and M.~Takeuchi,
  Phys.\ Rev.\  D {\bf 77}, 094012 (2008);
  T.~Kikuchi and N.~Okada,
  arXiv:0711.1506 [hep-ph];
  T.~Kikuchi and N.~Okada,
  Phys.\ Lett.\  B {\bf 661}, 360 (2008);
  A.~T.~Alan,
  arXiv:0711.3272 [hep-ph];
  A.~T.~Alan, N.~K.~Pak and A.~Senol,
  arXiv:0710.4239 [hep-ph];
  S.~L.~Chen, X.~G.~He, X.~P.~Hu and Y.~Liao,
  arXiv:0710.5129 [hep-ph];
  S.~L.~Chen, X.~G.~He, X.~Q.~Li, H.~C.~Tsai and Z.~T.~Wei,
  arXiv:0710.3663 [hep-ph];
  N.~G.~Deshpande, X.~G.~He and J.~Jiang,
  Phys.\ Lett.\  B {\bf 656}, 91 (2007);
  S.~L.~Chen, X.~G.~He and H.~C.~Tsai,
  JHEP {\bf 0711}, 010 (2007);
  S.~L.~Chen and X.~G.~He,
  Phys.\ Rev.\  D {\bf 76}, 091702 (2007);
  X.~G.~He and S.~Pakvasa,
  Phys.\ Lett.\  B {\bf 662}, 259 (2008);
  A.~Kobakhidze,
  Phys.\ Rev.\  D {\bf 76}, 097701 (2007);
  S.~Zhou,
  Phys.\ Lett.\  B {\bf 659}, 336 (2008);
  G.~J.~Ding and M.~L.~Yan;
  arXiv:0706.0325 [hep-ph].
  S.~L.~Chen, X.~G.~He and H.~C.~Tsai,
  JHEP {\bf 0711}, 010 (2007);
  C.~S.~Huang and X.~H.~Wu,
  Phys.\ Rev.\  D {\bf 77}, 075014 (2008);
  X.~Q.~Li, Y.~Liu and Z.~T.~Wei,
  arXiv:0707.2285 [hep-ph];
  M.~x.~Luo, W.~Wu and G.~h.~Zhu,
  Phys.\ Lett.\  B {\bf 659}, 349 (2008);
  G.~j.~Ding and M.~L.~Yan,
  arXiv:0709.3435 [hep-ph];
  G.~J.~Ding and M.~L.~Yan,
  Phys.\ Rev.\  D {\bf 77}, 014005 (2008);
  Y.~Liao,
  Phys.\ Rev.\  D {\bf 76}, 056006 (2007);
  Y.~Liao,
  arXiv:0708.3327 [hep-ph];
  Y.~f.~Wu and D.~X.~Zhang,
  arXiv:0712.3923 [hep-ph].


\bibitem{Dittmaier:1998dz}
  S.~Dittmaier, M.~Kramer, Y.~Liao, M.~Spira and P.~M.~Zerwas,
  Phys.\ Lett.\  B {\bf 441}, 383 (1998);
  W.~Bernreuther, M.~Fuecker and Z.~G.~Si,
  Phys.\ Rev.\  D {\bf 74}, 113005 (2006);
  A.~Brandenburg, Z.~G.~Si and P.~Uwer,
  Phys.\ Lett.\  B {\bf 539}, 235 (2002);
 W.~Beenakker, S.~Dittmaier, M.~Kramer, B.~Plumper, M.~Spira and P.~M.~Zerwas,
  Phys.\ Rev.\ Lett.\  {\bf 87}, 201805 (2001);
 W.~Bernreuther, A.~Brandenburg, Z.~G.~Si and P.~Uwer,
  Phys.\ Rev.\ Lett.\  {\bf 87}, 242002 (2001);
  Q.~H.~Cao, J.~Wudka and C.~P.~Yuan,
  Phys.\ Lett.\  B {\bf 658}, 50 (2007);
  C.~R.~Chen, F.~Larios and C.~P.~Yuan,
  Phys.\ Lett.\  B {\bf 631}, 126 (2005)
  [AIP Conf.\ Proc.\  {\bf 792}, 591 (2005)];
  W.~Bernreuther, A.~Brandenburg, Z.~G.~Si and P.~Uwer,
  Nucl.\ Phys.\  B {\bf 690}, 81 (2004);
  W.~Bernreuther, M.~Fuecker and Z.~G.~Si,
  Phys.\ Lett.\  B {\bf 633}, 54 (2006).

\bibitem{Choudhury:2007cq}
  D.~Choudhury and D.~K.~Ghosh,
  arXiv:0707.2074 [hep-ph];


\bibitem{Alan:2007ss}
  A.~T.~Alan and N.~K.~Pak,
  arXiv:0708.3802 [hep-ph].

\bibitem{Choi:1995kp}
  S.~Y.~Choi and K.~Hagiwara,
  Phys.\ Lett.\  B {\bf 359}, 369 (1995);
  B.~Grzadkowski, Z.~Hioki, K.~Ohkuma and J.~Wudka,
  Nucl.\ Phys.\  B {\bf 689}, 108 (2004).


\bibitem{Cheung:2007ap}
  K.~Cheung, W.~Y.~Keung and T.~C.~Yuan,
  Phys.\ Rev.\  D {\bf 76} (2007) 055003.

\bibitem{Brandenburg:2005uu}
  A.~Brandenburg and Z.~G.~Si,
  Phys.\ Lett.\  B {\bf 615}, 68 (2005).



\bibitem{Phinney:2007zz}
  N.~Phinney,
  ICFA Beam Dyn.\ Newslett.\  {\bf 42}, 7 (2007);
  J.~Brau, Y.~Okada and N.~Walker,
  arXiv:0712.1950 [physics.acc-ph];
  T.~Behnke {\it et al.},
  arXiv:0712.2356 [physics.ins-det].

\bibitem{Ginzburg:1982yr}
  I.~F.~Ginzburg, G.~L.~Kotkin, S.~L.~Panfil, V.~G.~Serbo and V.~I.~Telnov,
  Nucl.\ Instrum.\ Meth.\  A {\bf 219}, 5 (1984).


\bibitem{Grinstein:2008qk}
  B.~Grinstein, K.~Intriligator and I.~Z.~Rothstein,
  Phys.\ Lett.\  B {\bf 662}, 367 (2008).

\end{thebibliography}
\end{document}